\documentclass[aps,prl,reprint,superscriptaddress,twocolumn,showkeys,amsmath,amssymb,longbibliography,floatfix]{revtex4-2}
\usepackage[english]{babel}
\usepackage{amsmath,amssymb,bbm,mathrsfs,bm,braket,color,graphicx,comment,amsfonts,dsfont}
\usepackage[colorlinks,linkcolor=blue,citecolor=blue,urlcolor=blue]{hyperref}
\usepackage[mathscr]{euscript}
\usepackage{bm}

\newcommand{\pd}{{\phantom\dag}}
\begin{document}

\title{Non-Hermitian physics without gain or loss: the skin effect of reflected waves}

\author{Selma Franca}
\email{selma.franca@neel.cnrs.fr}
\affiliation{Institute for Theoretical Solid State Physics, IFW Dresden and W\"urzburg-Dresden Cluster of Excellence ct.qmat, Helmholtzstr. 20, 01069 Dresden, Germany}

\author{Viktor K{\"o}nye}
\affiliation{Institute for Theoretical Solid State Physics, IFW Dresden and W\"urzburg-Dresden Cluster of Excellence ct.qmat, Helmholtzstr. 20, 01069 Dresden, Germany}

\author{Fabian Hassler}
\affiliation{JARA-Institute for Quantum Information, RWTH Aachen University, 52056 Aachen, Germany}

\author{Jeroen van den Brink}
\affiliation{Institute for Theoretical Solid State Physics, IFW Dresden and W\"urzburg-Dresden Cluster of Excellence ct.qmat, Helmholtzstr. 20, 01069 Dresden, Germany}
\affiliation{Institute  for  Theoretical  Physics,  TU  Dresden,  01069  Dresden,  Germany}

\author{Cosma Fulga}
\affiliation{Institute for Theoretical Solid State Physics, IFW Dresden and W\"urzburg-Dresden Cluster of Excellence ct.qmat, Helmholtzstr. 20, 01069 Dresden, Germany}

\date{\today}
\begin{abstract}
Physically, one tends to think of non-Hermitian systems in terms of gain and loss: the decay or amplification of a mode is given by the imaginary part of its energy.
Here, we introduce an alternative avenue to the realm of non-Hermitian physics, which involves neither gain nor loss.
Instead, complex eigenvalues emerge from the amplitudes and phase-differences of waves backscattered from the boundary of insulators.
We show that for any strong topological insulator in a Wigner-Dyson class, the reflected waves are characterized by a reflection matrix exhibiting the non-Hermitian skin effect.
This leads to an unconventional Goos-H\"{a}nchen effect: due to non-Hermitian topology, waves undergo a lateral shift upon reflection, even at normal incidence.
Going beyond systems with gain and loss vastly expands the set of experimental platforms that can access non-Hermitian physics and show signatures associated to non-Hermitian topology.
\end{abstract}
\maketitle

\textit{\textcolor{blue}{Introduction}} --- Non-Hermitian physics describes a wide variety of quantum and classical systems \cite{Ashida2020, Bergholtz2021, Foa_Torres_2019}. 
In the quantum case, non-Hermitian operators model the coupling of systems to degrees of freedom that are outside of their Hilbert spaces, such as those of reservoirs or measurement devices \cite{Rotter2009}. 
In classical physics, non-Hermitian matrices commonly characterize optical systems \cite{El-Ganainy2018}, such as photonic crystals, but also mechanical \cite{Bertoldi2017} and acoustic metamaterials \cite{Cummer2016}, as well as electric circuits \cite{Weidemann2020, Helbig2020, Gupta2021}.

Despite their variety, these examples share gain and loss as their common physical origin of non-Hermiticity. 
Quantum systems may lose or gain quasiparticles when they are coupled to an external bath. 
In optics, the gain and loss of photons leads to complex-valued refractive indices, and thus to an effective description that makes use of non-Hermitian matrices \cite{Makris008}. 
Similarly, non-Hermiticity in mechanical systems and in electric circuits is due to the dissipation of energy produced through friction \cite{Yoshida2019}, and to the Joule heating caused by resistors. 
Thus, in all of these examples, the quantum or classical waves supported by the system are associated to complex eigenvalues. 
The real part encodes their excitation energy, or frequency, whereas the imaginary part describes their decay or amplification rate.

With this work, we offer a novel way of thinking about non-Hermiticity. 
Rather than ascribing a physical meaning to the real and imaginary parts of a complex eigenvalue, we imagine it to be made up of an amplitude and a phase.
To this end, we consider transport setups in which waveguides that support propagating modes are coupled to the boundaries of a topological insulator (TI) \cite{Hasan2010, Qi2011}. 
The waves which are backscattered from the TI boundary are characterized by the reflection matrix \cite{Nazarov2009}. 
This non-Hermitian operator has complex eigenvalues which describe how much of the wave is reflected (as opposed to transmitted), and what is the phase difference between the incident and the reflected wave.
Thus, within this framework, a zero-mode no longer has the meaning of a state at the Fermi level which does not decay but of a wave that is perfectly transmitted.

We prove that the reflection 
off all strong TIs in the Wigner-Dyson classes \cite{Dyson1962, Altland1997} (A, AI, or AII) exhibits a non-Hermitian skin effect (NHSE) \cite{Yao2018, MartinezAlvarez2018}.  
The novel way of thinking about non-Hermiticity thus
provides a dictionary that
directly relates the paradigmatic models introduced to study the NHSE to those of well-known strong TIs. 
The one-dimensional (1D) Hatano-Nelson model \cite{Hatano1996}, for example, is obtained as the reflection matrix from a 2D Chern insulator \cite{Haldane1988}. 
Similarly, the time-reversal invariant 1D and 2D NHSE \cite{Okuma2020} results when waves are backscattered from the boundaries of conventional 2D and 3D strong TIs \cite{Kane2005, Fu2007}.
As a result, non-Hermitian topology may be probed without introducing gain and loss into a system but rather by using the well-established tools of interferometry and (quantum) transport. 
Furthermore, this means that non-Hermitian topology leads to new experimental signatures, such as the lateral shift of a wave packet upon reflection from a TI edge.

We begin by briefly describing the topologically-protected non-Hermitian skin effect and showing that it is a universal property of reflection matrices from the boundaries of strong TIs in the Wigner-Dyson classes.
Afterwards, we explore the physical consequences of reflection matrix topology, focusing on a basic example: reflections from the boundary of a Chern insulator.

\textit{\textcolor{blue}{Non-Hermitian skin effect from reflections}} ---
The non-Hermitian skin effect is a manifestation of nontrivial topology \cite{Okuma2020, Borgnia2020, Lee2019a}: an extensive number of modes accumulates on the boundary of a system. 
Similar to conventional TIs such as Chern insulators, the boundary modes are present at all energies inside of a nontrivial bulk gap. 
Different from Hermitian systems, however, is that now the eigenvalues are complex, meaning that a `gap' is a 2D region of the complex plane (called a `point gap' \cite{Kawabata2019}), as opposed to a 1D interval on the real axis.

We consider strong TIs belonging to classes A, AI, or AII in the Altland-Zirnbauer classification \cite{Altland1997}. 
They occur in $D\geq 2$ space dimensions. 
These systems are probed in a conventional, two-terminal transport geometry: two semi-infinite waveguides which support propagating modes are attached to opposite boundaries of the TI. 
The reflection of waves incident on the TI boundary is described by a reflection matrix, $r$, which contains the probability amplitudes for any mode to be backscattered (Supplemental Material).

\begin{figure}[tb]
\includegraphics[width=8.6cm]{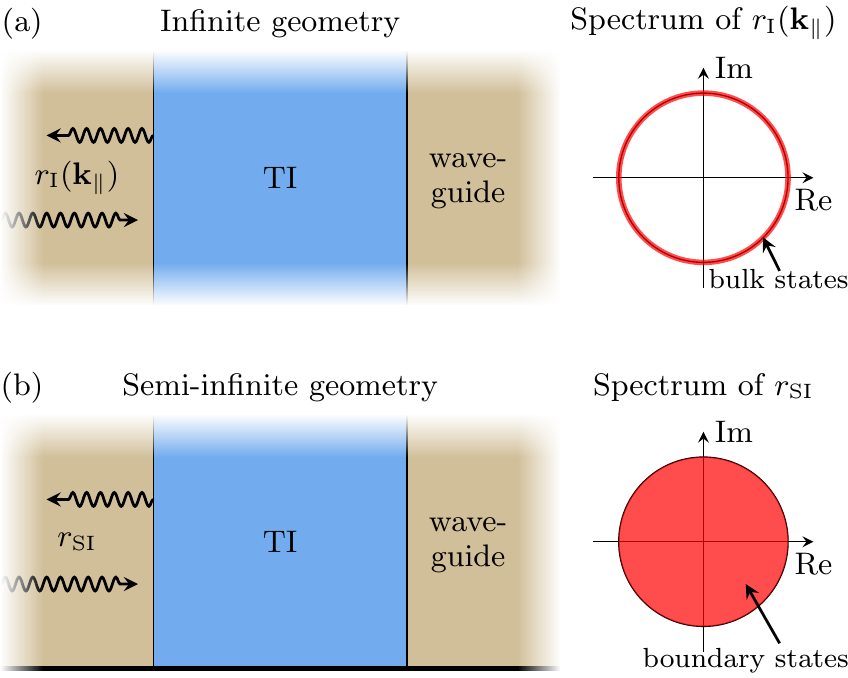}
\caption{
Transport geometries and spectrum of the reflection matrix. Panel (a) shows the infinite geometry and panel (b) shows the semi-infinite geometry.
In each panel, the spectrum of the reflection matrix is shown on the right. 
}
\label{fig:theorem}
\end{figure}

To distinguish between bulk and boundary effects, we consider two geometries. 
In the infinite geometry, both the $D$-dimensional TI and the two waveguides extend infinitely along the $(D-1)$ transversal directions, 
i.e. the directions parallel to the TI-waveguide interface [Fig.~\ref{fig:theorem}(a)]. 
We label the reflection matrix from the left waveguide as $r_{\rm I}({\bm k}_\parallel)$, where ${\bm k}_\parallel$ is a vector of the $(D-1)$ (conserved) momenta parallel to the interface. 
The semi-infinite geometry, with reflection matrix $r_{\rm SI}$, is obtained from the infinite one by introducing a single boundary along one of the transversal directions [Fig.~\ref{fig:theorem}(b)]. In both cases, we denote the eigenvalues of the reflection matrix by $z$.

The following theorem implies that the NHSE is a universal property of these reflection matrices.

{\bf Theorem:} Let $H_D$ be a $(D>1)$-dimensional strong topological insulator in one of the classes A, AI, or AII. 
In the two-terminal geometries described above, the reflection matrix from the boundary of $H_D$ shows the non-Hermitian skin effect. This means that it exhibits the following properties:

\noindent
($i$) In the infinite geometry, the spectrum of $r_{\rm I}$ is identical to the unit circle in the complex plane, $|z|=1$. 
Therefore, $r_{\rm I}$ has two distinct point gaps: one inside the unit circle, $|z|<1$, and one outside the unit circle, $|z|>1$ [see Fig.~\ref{fig:theorem}(a)].

\noindent
($ii$) In the semi-infinite geometry, the spectrum of $r_{\rm SI}$ covers the entire unit disk, $|z|\leq 1$, which means that the point gap at $z=0$ closes as soon as a boundary is introduced (in the transversal direction). 
For any value of $z$ inside the unit disk, $r_{\rm SI}$ has at least one eigenstate localized at the boundary of the waveguide  [see Fig.~\ref{fig:theorem}(b)].

The idea for the proof of this Theorem is briefly outlined below. The full proof is given in Supplemental Material~\cite{sm}.
There, we also discuss how the Theorem can potentially be extended to topological superconductors \cite{Qi2011}, weak topological insulators \cite{Fu2007}, topological crystalline insulators \cite{Fu2011}, and higher-order topological insulators \cite{Benalcazar2017, Schindler2018}.

To show the presence of the NHSE, we introduce the auxiliary Hamiltonian
\begin{equation}\label{eq:Hdm1}
 H_{D-1}(z) =
 \begin{pmatrix}
  0 & r^\pd_{\rm I}({\bm k}_\parallel) - z \\
  r^\dag_{\rm I}({\bm k}_\parallel) - z^* & 0 
 \end{pmatrix},
\end{equation}
which depends on the complex parameter $z$.
This Hamiltonian is Hermitian and $(D-1)$-dimensional, since it depends on the momenta ${\bm k}_\parallel$. 
Moreover, it has chiral symmetry, $\tau_z H_{D-1}(z) \tau_z = - H_{D-1}(z)$, with $\tau_z$ the Pauli matrix acting in the $2\times 2$ grading of Eq.~\eqref{eq:Hdm1}.
For a TI in class A, AI, or AII, $H_{D-1}(z)$ belongs to class AIII, BDI, or DIII, respectively.

Note that for any eigenstate of $H_{D-1}(z)$ at zero energy, there is an eigenstate of $r^\pd_{\rm I}$ with complex eigenvalue $z$ (Supplemental Material). 
This enables us to derive the spectrum of the reflection matrix by examining the gap closings of $H_{D-1}$. 
The latter is a strong topological insulator protected by chiral symmetry when $z=0$ \cite{Fulga2012, Schulz-Baldes2020}, and becomes a trivial, atomic insulator for $z\to\infty$. 
Therefore, for any path in the complex plane from $z=0$ to $\infty$, a topological phase transition must occur in $H_{D-1}(z)$, signaled by a closing of its bulk gap. 
This means that the spectrum of $r_{\rm I}$ forms a closed loop encircling the origin of the complex plane. 
Deep in the insulating regime, the waves cannot be transmitted across the TI with a gapped bulk. Thus, $r_{\rm I}$ is unitary with the spectrum on the unit circle. 
The closed loop formed by its spectrum must therefore be the full unit circle in the complex plane.

To show the existence of boundary states in the semi-infinite geometry, we simply replace $r_{\rm I}$ with $r_{\rm SI}$ in Eq.~\eqref{eq:Hdm1}. 
$H_{D-1}(z)$ is still a strong TI for any $|z|<1$, as before, but now with a boundary and the corresponding boundary mode(s) at zero energy. As a result, the reflection matrix $r_{\rm SI}$ will have at least one boundary state for any $z$ inside the unit disk 
and the spectrum of $r_{\rm SI}$ covers the entire unit disk.

\begin{figure*}[tb]
\includegraphics[width=16.44cm]{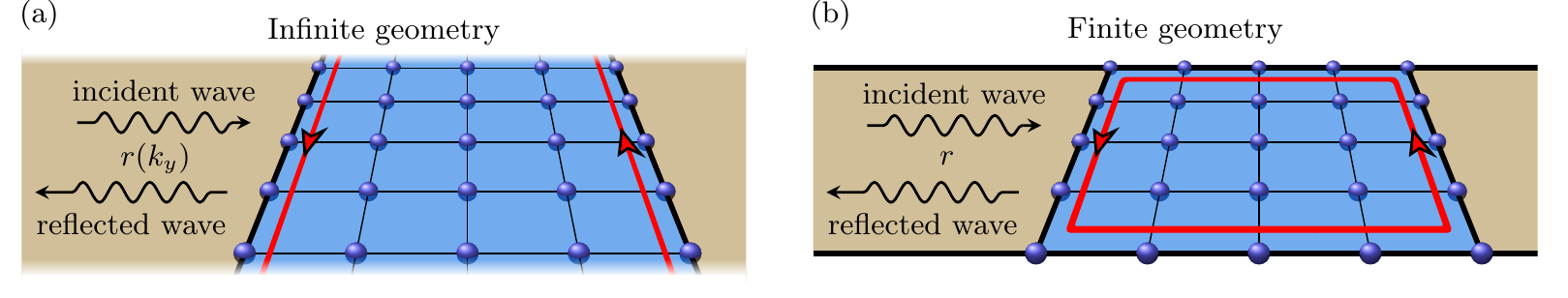}
\caption{
Scattering setup.
The Chern insulator (blue) is connected to two waveguides (beige). 
In panel (a), the system is infinite in the $y$ (vertical) direction, and the reflection matrix depends on $k_y$. 
In panel (b), both the Chern insulator and the waveguides have a finite extent in the transversal direction.
}
\label{fig:setup}
\end{figure*}

\textit{\textcolor{blue}{Reflections from a Chern insulator}} -- 
For concreteness, we examine the physical consequences of non-Hermitian topology of the reflection matrices by focusing on a basic example. 
We take $H_{D=2}$ to be a Chern insulator, using the toy model introduced by Qi, Wu, and Zhang (QWZ) \cite{Qi2006}. 
The two-band, square-lattice model has the form 
$H(\bm{k}) = \bm{d}(\bm{k}) \cdot {\bm{\sigma}}$, 
with  the momentum
${\bm k}=(k_x,k_y)$, 
the vector of Pauli matrices
$\bm{\sigma}=(\sigma_x, \sigma_y, \sigma_z)$, 
and 
${\bm d}({\bm k})=(\sin k_x, \sin k_y, m-\cos k_x-\cos k_y)$. 
The system is gapped and topologically nontrivial for $0<|m|<2$, with a single chiral edge mode localized on its boundary. 
The bulk gap closes at $|m|=2$, signaling a transition to a trivial insulating phase at $|m|>2$.

We form the infinite geometry by connecting waveguides to an infinite ribbon of the QWZ model, with a thickness of $L$ sites in the $x$ direction [Fig.~\ref{fig:setup}(a)].
The waveguide (or lead) Hamiltonian,
$H_l = -2\cos k_x \sigma_0$, 
is time-reversal symmetric and consists of independent, decoupled chains, each of which probes one site of the system boundary.
As such, the waveguide supports both incoming as well as outgoing modes for every value of the transversal momentum $k_y$.

For $0<m<2$, the presence of a single chiral edge mode at the system boundary is a consequence of the nonzero Chern numbers of the bulk bands. 
The Chern number of the lower band is given by the winding number of the reflection matrix \cite{Braeunlich2009, Fulga2012},
\begin{equation}\label{eq:ci_inv}
C = \frac{1}{2\pi i} \int_0^{2\pi}  d k_y \frac{d}{d k_y} \text{log} \det{r(k_y)},
\end{equation}
and takes on the value $C=1$.

\begin{figure}[tb]
\includegraphics[width=\columnwidth]{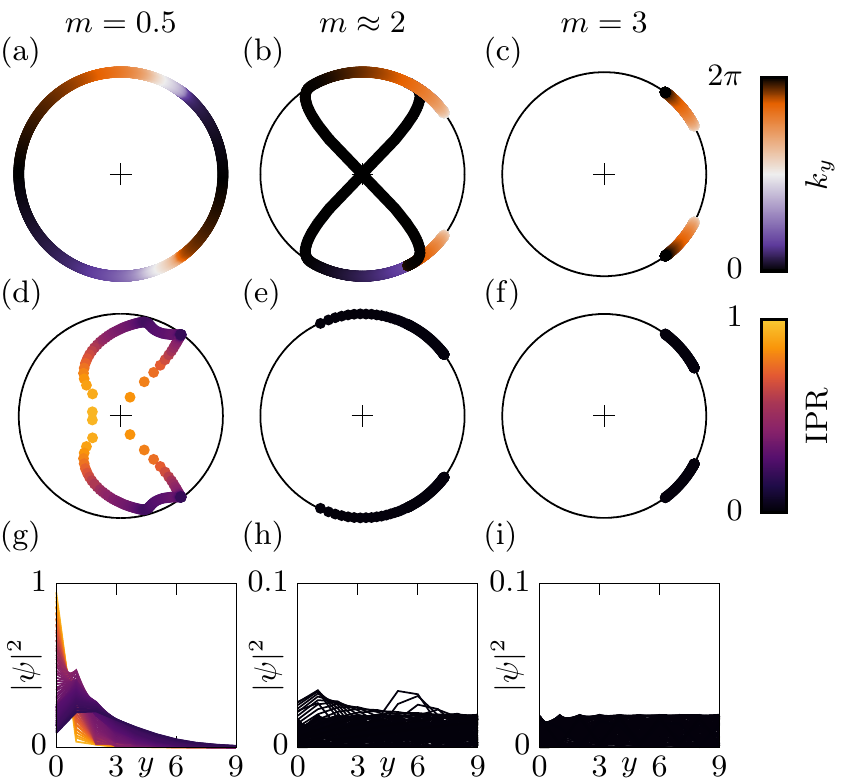}
\caption{
Spectra and eigenvectors of the reflection matrix. 
The origin of the complex plane, $z = 0$, is indicated with a cross. 
In panels (a--c), the spectra are plotted for a QWZ ribbon (100 sites long), as shown in Fig.~\ref{fig:setup}(a). 
There reflection matrix eigenvalues form two `bands', corresponding to the two orbitals per site of the QWZ model.
The color scale indicates the value of momentum $k_y$ at which $r(k_y)$ is calculated. 
In panels (d--f), the spectra are plotted for a finite system ($100\times 100$ sites), as in Fig.~\ref{fig:setup}(b). 
The color scale indicates the inverse participation ratio (IPR) of the eigenstates defined as $\textrm{IPR}_i = \sum_{y = 0}^{L-1} |\psi_i(y)|^4$. 
Panels (g--i) show the real space probability distribution of all reflection matrix eigenstates corresponding to the eigenvalues in panels (d--f). 
Here, $y=0$ denotes the bottom boundary of the waveguide (only the first 10 sites are shown). 
}
\label{fig:phasetrans}
\end{figure}

The above relation, originally a consequence of Laughlin's pumping argument, can now be understood as a manifestation of the non-Hermitian topology of the reflection matrix. 
Eq.~\eqref{eq:ci_inv} is the (nonzero) winding number of the determinant of $r$ and thus 
the eigenvalues of the unitary reflection matrix wind, as a function of $k_y$, around the unit circle in the complex plane [Fig.~\ref{fig:phasetrans}(a)]. This matches with the statement of Theorem ($i$).

The Chern number of the QWZ model induces a topologically-nontrivial NHSE in its reflection matrix. 
The latter thus becomes topologically equivalent to the Hatano-Nelson model, with Eq.~\eqref{eq:ci_inv} the invariant of the Hatano-Nelson model \cite{Okuma2020}.
As a result, for a finite transversal extent, all the eigenstates of $r$ are localized at one boundary of the waveguide [Fig.~\ref{fig:phasetrans}(g)],
consistent with the statement of Theorem ($ii$).
When the Hamiltonian undergoes a topological-to-trivial transition, for instance by changing $m$ from $\tfrac12$ to 3, the $z=0$ point gap of the reflection matrix closes and reopens [Fig.~\ref{fig:phasetrans}]. 
For $m=3$, both the bulk gap of the QWZ model and the point gap of $r$ are trivial, $C=0$, and no skin effect occurs.

Physically, the reflection matrix converts incoming states into outgoing states within the same waveguide. 
Thus, an eigenstate of $r$ corresponds to a wave which is backscattered without any change in its spatial profile. 
This provides an experimental transport signature associated to the non-Hermitian topology of the reflection matrix. 
If an incoming wave is an eigenstate of $r$ with a spatial profile as those shown in Fig.~\ref{fig:phasetrans}(g), which are localized on the waveguide boundary due to the NHSE, then the reflected wave will have the same profile. 
Such profiles might be obtained, for instance, by using beam-forming techniques as in Refs.~\cite{Valagiannopoulos2017, Valagiannopoulos2018, Valagiannopoulos2019}.
The complex eigenvalue associated to this eigenstate, $z=\rho e^{i \phi}$ with $\rho\ge0$, determines the ratio between the overall amplitudes of the reflected and incoming wave ($\rho$), as well as the phase difference between them ($\phi$).

In the Hatano-Nelson model, the presence of the NHSE has an intuitive explanation: hoppings are nonreciprocal. 
If, say, the hopping to the left is stronger than the hopping to the right, then all bulk states will be pumped leftwards until they eventually accumulate on the system boundary. 
Thus, the NHSE is associated with a unidirectional persistent current in the bulk of the system \cite{Lee2019, Zhang2020}. 
When the non-Hermitian Hamiltonian acts on an initial state, it will produce a time-evolved state which is unidirectionally shifted due to this current.

Since the reflection matrix is topologically equivalent to the Hatano-Nelson model, the same shift occurs when it acts on an incoming wave, resulting in an outgoing wave at a different position.
Heuristically, this can be understood by noting that when an incoming mode reaches the Chern insulator boundary, it will couple to its topological edge state. 
Due to the unidirectional nature of this edge state, the reflected wave will be laterally shifted in the direction of propagation of the  edge mode.
This is a second experimental signature of the non-Hermitian topology of $r$: the nonreciprocal Goos-H\"{a}nchen effect \cite{Ma2020}. 

\begin{figure}[tb]
\includegraphics[width = 8.6cm]{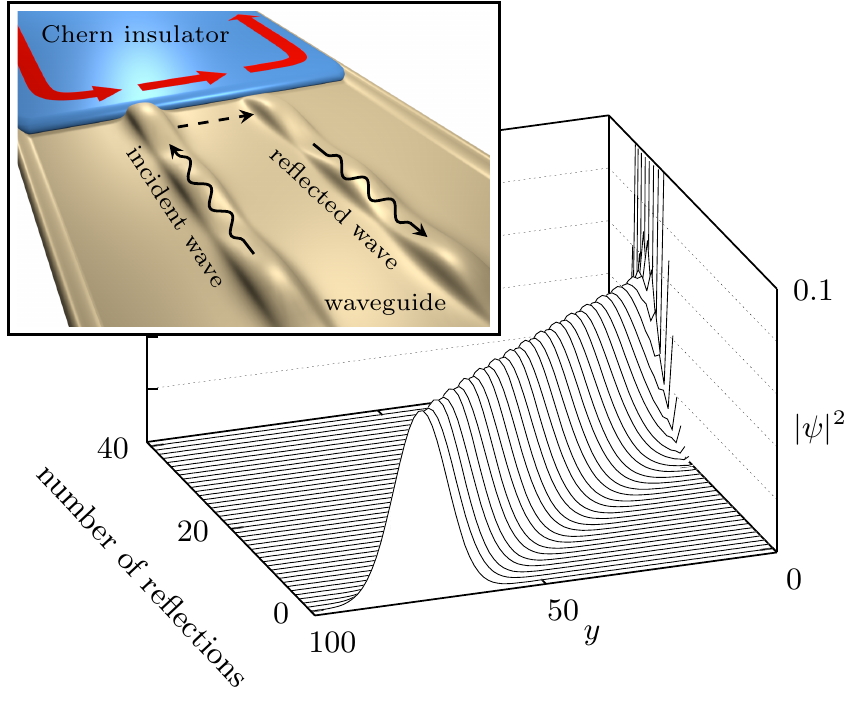}
\caption{
The nonreciprocal Goos-H\"{a}nchen effect of a wave packet reflected multiple times from a Chern insulator boundary. 
The inset shows the effect schematically: the incident wave packet is shifted laterally upon reflection. The result of the numerical simulation is shown as a waterfall plot of the real space probability distribution of a propagating mode along the direction parallel to the Chern insulator boundary, $y$. 
The first curve is the initial wave packet shape (normalized), whereas subsequent curves show the wave packet after multiple reflections from the TI boundary. 
We used a QWZ model of $100\times100$ sites, setting $m=\tfrac12$.
}
\label{fig:GHeffect}
\end{figure}

We examine the lateral shift of reflected waves in Fig.~\ref{fig:GHeffect}, which shows a wave packet of a propagating mode initially localized in the left side of the waveguide. 
The mode is trapped between the Chern insulator boundary and a perfect mirror, and thus undergoes multiple reflections. 
After each reflection from the TI boundary, the lateral position of the propagating mode is shifted in the  direction of motion of the edge state, until eventually reaching the waveguide boundary. 
At that point the mode is transmitted into the other lead by means of the chiral edge state connecting them.

While phenomenologically similar, this lateral shift is distinct from the conventional Goos-H\"{a}nchen as well as the Imbert-Fedorov effect \cite{Bliokh2013}. 
In the latter cases, the shift of the incoming beam only occurs for waves impinging at an angle. 
Here, the lateral shift is present even for normal incidence. The unidirectional shift is due to the non-Hermitian topology of the reflection matrix and does not occur if $r$ is not topological.

\textit{\textcolor{blue}{Experimental realizations}} -- 
The experimental signatures we have introduced above can be readily tested using available experimental platforms, by performing transport measurements and interferometry. 
In fact, some of these experiments have already been performed, though their connection to non-Hermitian topology has been overlooked. 
The winding of the eigenvalues has been first measured seven years ago, in a microwave network \cite{Hu2015}. 
Last year, the nonreciprocal Goos-H\"{a}nchen effect has been observed in a photonic crystal \cite{Ma2020}. 
Our Theorem shows that these observations are actually connected to the non-Hermitian skin effect, which is a universal property of reflected waves from TIs.
We predict that unconventional Goos-H\"{a}nchen effects will occur when waves are reflected from any TI boundaries. 
For instance, when backscattered from a quantum spin-Hall edge, the incoming beam will experience a spin-dependent lateral shift.

There are a number of advantages to performing experiments on non-Hermitian topology by using reflection matrices. 
A conventional, Hermitian topological insulator can be used instead of a system where gain and loss need to be introduced in a controlled way. 
Related to this, reflection matrix experiments eliminate the problem of dynamical instability, which occurs whenever overall gain is stronger than loss. 
This is particularly important for non-Hermitian topology which requires balanced gain and loss, such as the so-called `PT-symmetric' systems \cite{El-Ganainy2018}. 
There, balancing the amplification and attenuation of waves requires fine-tuning the system on the boundary of its dynamically-unstable regime \cite{Schindler2011}. 
In contrast, reflection matrices are stable regardless of whether their eigenvalues are above or below the real axis. 
This is because the eigenvalues encode phase differences between outgoing and incoming modes.

\textit{\textcolor{blue}{Conclusion}} -- 
We have shown that non-Hermitian physics does not require gain and loss but can arise when waves are backscattered from the boundary of an insulator. 
Apart from a different realization of non-Hermitian physics, our novel viewpoint leads to the insight that, when the insulator is a strong TI in a Wigner-Dyson class, reflected waves universally show a non-Hermitian skin effect. 
Moreover, we predict an unconventional, nonreciprocal Goos-H\"{a}nchen effect, in which non-Hermitian topology causes incoming, normal-incidence waves to undergo a lateral shift upon reflection.

\textit{\textcolor{blue}{Acknowledgments}} --
We thank Ulrike Nitzsche for technical assistance. 
This work was supported by the Deutsche Forschungsgemeinschaft (DFG, German Research Foundation) under Germany's Excellence Strategy through the W\"{u}rzburg-Dresden Cluster of Excellence on Complexity and Topology in Quantum Matter -- \emph{ct.qmat} (EXC 2147, project-id 390858490) and under Germany's Excellence Strategy -- Cluster of Excellence Matter and Light for Quantum Computing (ML4Q) EXC 2004/1 -- 390534769.

\textit{\textcolor{blue}{Author Contributions}} --
CF initiated and oversaw the project. 
SF and VK performed numerical calculations and produced the figures. 
All authors contributed towards formulating and proving the Theorem, interpreting the results, and writing the manuscript.

\bibliography{NHr}

\end{document}


\title{Supplemental Material to: Non-Hermitian physics without gain or loss: the skin effect of reflected waves}

\author{Selma Franca}
\email{selma.franca@neel.cnrs.fr}
\affiliation{Institute for Theoretical Solid State Physics, IFW Dresden and W\"urzburg-Dresden Cluster of Excellence ct.qmat, Helmholtzstr. 20, 01069 Dresden, Germany}

\author{Viktor K{\"o}nye}
\affiliation{Institute for Theoretical Solid State Physics, IFW Dresden and W\"urzburg-Dresden Cluster of Excellence ct.qmat, Helmholtzstr. 20, 01069 Dresden, Germany}

\author{Fabian Hassler}
\affiliation{JARA-Institute for Quantum Information, RWTH Aachen University, 52056 Aachen, Germany}

\author{Jeroen van den Brink}
\affiliation{Institute for Theoretical Solid State Physics, IFW Dresden and W\"urzburg-Dresden Cluster of Excellence ct.qmat, Helmholtzstr. 20, 01069 Dresden, Germany}
\affiliation{Institute  for  Theoretical  Physics,  TU  Dresden,  01069  Dresden,  Germany}

\author{Cosma Fulga}
\affiliation{Institute for Theoretical Solid State Physics, IFW Dresden and W\"urzburg-Dresden Cluster of Excellence ct.qmat, Helmholtzstr. 20, 01069 Dresden, Germany}

\begin{abstract}
In this Supplemental Material, we describe in more details how the scattering matrix is calculated and its symmetry constraints. We then provide additional details on the proof presented in the main text, and discuss the extension to other symmetry classes. In addition, we show how the finite size effects affect the non-Hermitian skin effect of the reflection matrix.  Lastly, we discuss how the nonreciprocal Goos-H\"anchen shift depends on the parameters of the Qi-Wu-Zhang model studied in the main text.
\end{abstract}

\date{\today}
\maketitle

\section{Scattering matrix and its symmetries}
 
We determine the scattering matrix of the system by solving the tight-binding equations of the full, system-plus-waveguide Hamiltonian. This essentially consists of matching the wavefunctions of the propagating lead modes to those of the central scattering region. The procedure is explained in the Kwant paper, Ref.~\cite{Groth2014}, whose code we use to produce all of our numerical results. Our codes are available at \cite{code}.

The scattering matrix is unitary, taking the form:
\begin{equation} \label{eq:smatrix}
 S=\begin{pmatrix}
         r & t' \\
         t & r' \\
        \end{pmatrix},
\end{equation}
where the transmission blocks $t^{(\prime)}$ contain the probability amplitudes for waves to be transmitted from one waveguide to the other, whereas the reflection blocks $r^{(\prime)}$ contain the probability amplitudes for states to be backscattered. 
Throughout, we have considered the scattering matrix computed at the Fermi level, which we set to zero energy.

When the scattering region and the waveguides obey time-reversal (TRS), particle-hole (PHS), or chiral symmetries, so does the scattering matrix. 
For future use, we list how these symmetries constrain the form of $S$. 
The detailed derivation can be found in Refs.~\cite{Fulga2012, Franca2021}, so here we only state the results.

Both TRS (${\cal T}$) and PHS (${\cal P}$) are anti-unitary, whereas chiral symmetry (${\cal C}$) is unitary. 
All three are local, meaning that they commute with spatial translations. 
They take the form 
${\cal P} = U_{\cal P} {\cal K}$, ${\cal T} = U_{\cal T} {\cal K}$, and ${\cal C}=U_{\cal C}$,
respectively, with ${\cal K}$ denoting complex conjugation and $U_{{\cal T}, {\cal P}, {\cal C}}$ unitary operators. 
When the full, system-plus-waveguide Hamiltonian respects either of these symmetries,
\begin{equation}\label{eq:sym_fullH}
 H = U_{\cal T}^\pd H^* U_{\cal T}^\dag, \quad H = -U_{\cal P}^\pd H^* U_{\cal P}^\dag, \quad H = -U_{\cal C}^\pd H U_{\cal C}^\dag,
\end{equation}
then the scattering matrix obeys:
\begin{equation}\label{eq:sym_S}
 U_{\cal T}^\pd S^* U_{\cal T}^\dag = S^\dag, \quad U_{\cal P}^\pd S^* U_{\cal P}^\dag = S, \quad \quad U_{\cal C}^\pd S^\dag U_{\cal C}^\dag = S.
\end{equation}
Since the reflection matrix is a diagonal sub-block of S, the same equations apply to $r$.

The above equations express symmetries of the full tight-binding model, which includes both the waveguides and the central region. 
Thus, if the waveguides obey TRS but the system Hamiltonian breaks this symmetry, as happens for the QWZ model discussed in the main text, then the scattering matrix will not respect TRS. 
Also, note that Eqs.~\eqref{eq:sym_S} treat outgoing and incoming modes identically, since the same unitary operators appear on both sides of $S$. 
This is a `natural basis', meaning that the degrees of freedom of propagating modes (e.g. position, spin, sublattice, electron-hole) have the same meaning regardless of the direction of propagation of the mode.

\section{Proof of the Theorem}

We begin by focusing on class A, followed by classes AI and AII. 
Afterwards, we will comment on ways this statement may be extended to cover different types of gapped topological phases.

We begin with the infinite geometry, in which we take the class A reflection matrix $r_{\rm I}$ to be unitary. 
This is justified due to the fact that the bulk of the strong TI is gapped, meaning that the transmission between waveguides is exponentially suppressed. 
As explained in the main text, the auxiliary Hamiltonian $H_{D-1}(z)$ [Eq.~(1) of the main text] obeys chiral symmetry, 
$\tau_z H_{D-1}(z) \tau_z = - H_{D-1}(z)$. 
Since the reflection matrix describes a system in class A, no other symmetries constrain its form, and chiral symmetry is the only symmetry of $H_{D-1}(z)$. 
Therefore, this $(D-1)$-dimensional Hamiltonian belongs to class AIII in the Altland-Zirnbauer classification.

Throughout the following, we shall use a simple observation relating the spectra of the dimensionally-reduced Hamiltonian and that of the reflection matrix. 
If $H_{D-1}(z)$ has a zero-energy eigenstate, then the reflection matrix has an eigenstate with eigenvalue $z$. 
Thus:
\begin{equation}\label{eq:lemma}
 0 \in \sigma \left[ H_{D-1}(z) \right] \Leftrightarrow z \in \sigma \left[ r_{\rm I} \right],
\end{equation}
where $\sigma [\ldots]$ denotes the spectrum. The right implication follows immediately from
$\det [H_{D-1}(z)] = |\det [r_{\rm I} - z]|^2 =0$.
Conversely, for the left implication,
$r^\pd_{\rm I}\ket{\psi_B}= z \ket{\psi_B}$ 
means that 
$(0, \ket{\psi_B})^T$ 
is a zero-energy eigenstate of $H_{D-1}(z)$.

The above relation means that to determine the spectrum of the reflection matrix, it is sufficient to know what are the values of $z$ for which the gap of $H_{D-1}(z)$ closes. 
This can be done by examining which topological phases are realized by the latter.

We use a result obtained in Refs.~\cite{Fulga2012, Schulz-Baldes2020}, namely that if $H_D$ is a strong topological insulator in class A, then $H_{D-1}(z=0)$ must be a strong topological insulator in class AIII (protected by chiral symmetry). 
Furthermore, we also know that for large $z$, $H_{D-1}(z)$ is a trivial system in class AIII: $\lim_{|z|\to\infty} H_{D-1}(z) = -({\rm Re}\,z) \tau_x + ({\rm Im}\,z) \tau_y$ is an atomic insulator. 
Since the Hamiltonian obeys chiral symmetry for any $z$, a topological phase transition must occur when changing $z=0$ to $z\to\infty$ along any path in the complex plane $({\rm Re}\,z, {\rm Im}\,z)$. 
Therefore, there must exist a closed contour of $z$, which surrounds $z=0$, for which the bulk gap of $H_{D-1}$ closes. 
By using Eq.~\eqref{eq:lemma}, the spectrum of $r_{\rm I}$ forms a closed contour around $z=0$, which must be the unit circle because the reflection matrix is unitary. 
This concludes the proof of the first part of our Theorem.
Note that we have determined the spectrum of the reflection matrix in a model independent way.
Thus, regardless of the number of eigenvalues of $r_{\rm I}$ or their specific dependence on ${\bm k}_\parallel$, they must cover the full unit circle.

The proof of the second part of the Theorem is completely analogous but it refers to the semi-infinite geometry instead of the infinite one. 
We form a $D-1$ dimensional Hamiltonian, just as in Eq.~(1) of the main text, but now using $r_{\rm SI}$ instead of $r_{\rm I}$. 
This Hamiltonian belongs to class AIII for any $z$, as before, and is still nontrivial when $z=0$. 
Since it has a boundary, this means it will have at least one zero energy boundary state for all values of $z$ for which it is nontrivial, that is, everywhere inside the unit disk. 
By Eq.~\eqref{eq:lemma}, the spectrum of $r_{\rm SI}$ must cover the entire unit disk, $|z|<1$. 
Furthermore, since edge states of $H_{D-1}$ correspond to edge states of $r_{\rm SI}$, there will exist (at least one) reflection matrix eigenstate localized at the waveguide boundary for every $|z|<1$.

Compared to class A, in classes AI and AII the main difference comes from the fact that the lower-dimensional Hamiltonian $H_{D-1}(z)$ forms a topological phase which is not protected by chiral symmetry alone but also by time-reversal symmetry.
We will use the same approach of reflection matrix-based dimensional reduction as before, that of Refs.~\cite{Fulga2012, Schulz-Baldes2020}, but expressed in the basis of Eqs.~\eqref{eq:sym_S}.
With TRS, the reflection matrix obeys
\begin{equation}\label{eq:rtrs}
 U_{\cal T}^\pd r^*_{\rm I}({\bm k}_\parallel) U_{\cal T}^\dag = r^\dag_{\rm I}(-{\bm k}_\parallel),
\end{equation}
where we have focused on the infinite geometry in order to show the transformation of momentum induced by the TRS. 
Using Eq.~\eqref{eq:rtrs} in Eq.~(1) of the main text, we conclude that the latter is time-reversal symmetric for any $z$, with a new TRS operator 
$\widetilde{\cal T} = U^\dag_{\cal T}\otimes \tau_x {\cal K}$. 
$\tau_x$ acts on the $2\times 2$ grading characterizing the blocks of $H_{D-1}(z)$, whereas $U^\dag_{\cal T}$ acts within each block. 
Since, as before, the lower-dimensional Hamiltonian has a chiral symmetry $\tau_z$, we conclude it must also posses the product of TRS and chiral symmetry, which is a PHS 
$\widetilde{\cal P} = U^\dag_{\cal T}\otimes \tau_y {\cal K}$. 
Thus, class AI maps to CI under dimensional reduction, and class AII maps to DIII, consistent with Ref.~\cite{Fulga2012}.

The rest of the proof is identical to that of class A. 
The lower-dimensional Hamiltonian is nontrivial at $z=0$ and trivial for $z\to\infty$, changing $z$ does not break the protecting symmetries, which means that the spectrum of the reflection matrix must be the unit circle in the infinite geometry and the unit disk in the semi-infinite geometry.

\section{Extension to other topological phases}

The Altland-Zirnbauer classification consists of ten classes, and above we have only focused on three of them: the Wigner-Dyson classes A, AI, and AII. 
There are two reasons for this. 
First, in five of the remaining symmetry classes, namely the chiral ones, we do not know how to show that $H_{D-1}(z=0)$ must be a strong TI, so that the above reasoning cannot be applied. 
Second, in class C and class D, while we do know that $H_{D-1}(z=0)$ is a strong TI, there is no guarantee that this will remain the case away from $z=0$.

In class C and class D, our proof does not work because the dimensionally-reduced Hamiltonian we construct, $H_{D-1}(z)$, breaks the symmetries responsible for its topological protection away from $z=0$. 
In both of these classes, the system obeys PHS, 
${\cal P} = U_{\cal P} {\cal K}$, 
so that the reflection matrix respects 
$U_{\cal P}^\pd r^* U_{\cal P}^\dag = r$. 
Plugging this relation into Eq.~(1) of the main text, we get
\begin{equation}\label{eq:badphs}
 [H_{D-1}(z)]^* = - (\tau_z U_{\cal P})^\dag H_{D-1}(z^*) (\tau_z U_{\cal P}).
\end{equation}
Equation~\eqref{eq:badphs} is a valid particle-hole symmetry only as long as $z=z^*$. 
As before, Ref.~\cite{Fulga2012} has shown that $H_{D-1}$ is nontrivial at $z=0$, being protected by PHS. 
For $z\to\infty$, it is still an atomic insulator. 
However, now there will exist paths between these two points for which particle-hole symmetry is broken, and thus the bulk gap closing of $H_{D-1}$ could be avoided.

Just because our proof does not apply to classes C and D does not mean that $r$ is forbidden from hosting a NHSE. 
In fact, it clearly does when waves are backscattered from the boundary of a 2D strong topological phase in these classes (e.g. the $p+ip$ topological superconductor in class D). 
This is because the invariant responsible for edge modes in 2D class C and D is still the winding number of $\det\, r$, so the reflection matrix is again topologically equivalent to the Hatano-Nelson model. 
We conjecture that $r$ will still have the NHSE in class C and D whenever the original $H_D$ has a $\mathbb{Z}$ topological classification. 

Finally, we believe that it should be possible to show that $r$ has a NHSE also in other types of gapped topological phases, such as weak topological insulators, topological crystalline insulators, and higher-order topological insulators (HOTI). 
These phases typically host gapless modes on some but not all boundaries. 
Therefore, our construction of the semi-infinite geometry would have to take this into account, by selecting one (or more for HOTIs) boundary directions in order to ensure that the two waveguides are connected by a conducting boundary state. 
Otherwise, the reflection matrix would remain unitary regardless of whether the boundary is present or not, which means that the point gap at $z=0$ would not close. 
We note that a reflection matrix-based dimensional reduction has been already developed for HOTIs in Ref.~\cite{Geier2018}, and it seems possible to use it in a similar way to our proof above.

\section{Finite size effects}

Through our Theorem, we have related the non-Hermitian skin effect present in reflection matrices of strong TIs in classes A, AI and AII to the nontrivial winding of their eigenvalues in momentum space. 
For a Chern insulator, the results shown in Fig.~2 of the main text corroborate this connection for the nontrivial phase ($m = \tfrac12$). 
However, at the phase transition point ($m \approx 2$) [Fig.~3(b) of the main text] we see that there are two separate windings of the reflection matrix eigenvalues. 
These windings occur around some finite $z$ because $r(k_y)$ is singular at the phase transition point due to the bulk gap closing. 
What is at odds with our Theorem is that we observe no NHSE when the boundaries are introduced, see Fig.~3(h) of the main text. 
Here, we argue that this is due to large finite size effects.

\begin{figure}[h!]
\includegraphics[width=\columnwidth]{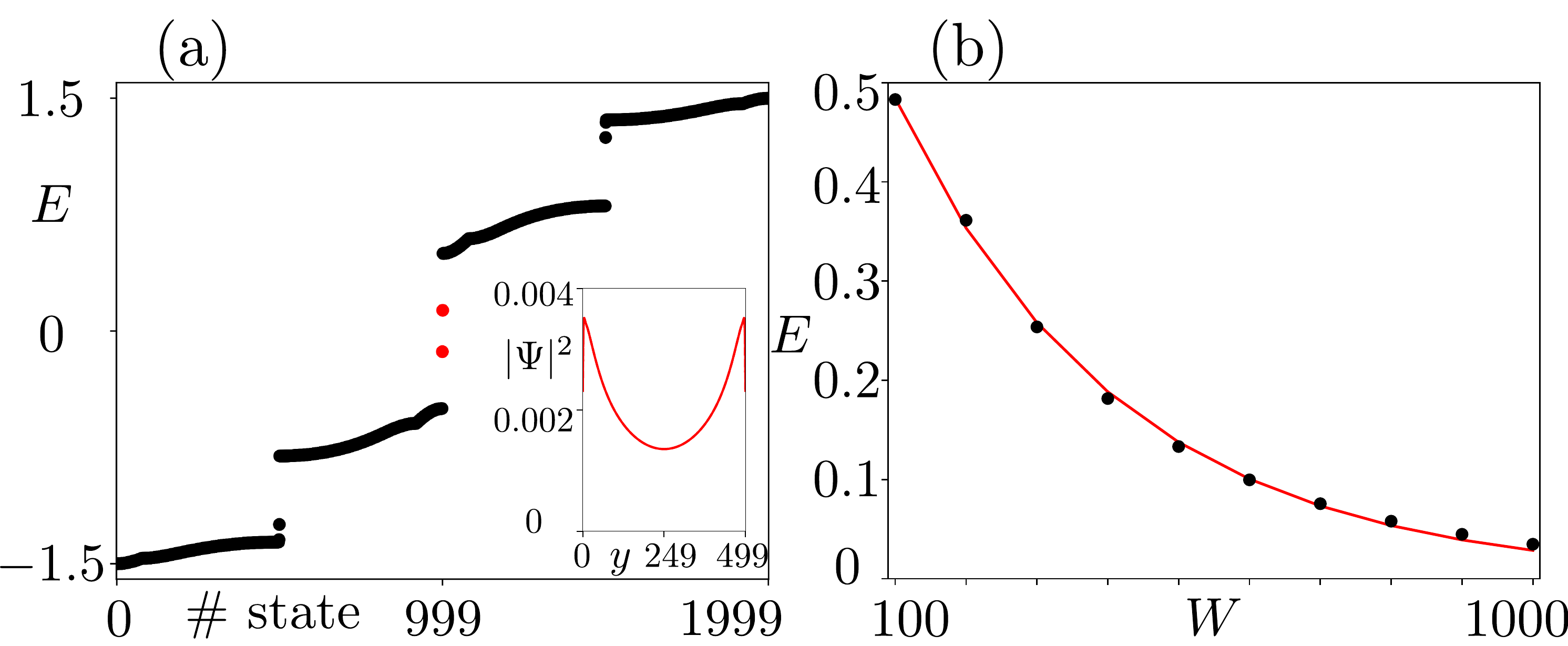}
\caption{Finite-size energy splitting.
Panel (a) shows the spectrum of the doubled Hamiltonian obtained for a scattering region that is $100$ ($500$) sites long (wide), using the parameters of Fig.~3(e) of the main text. 
The inset is the probability distribution of lowest-lying energy states. 
Panel (b) depicts how the energy of these modes depends on the width $W$ of the scattering region. 
The red curve represents the exponential fit 
yielding a localization length of $\xi\simeq 318$ sites.
The other parameters remain unchanged compared to panel (a). }
\label{fig:finite_size}
\end{figure}

To that end, we construct the doubled Hamiltonian $H_{D-1}(z)$ using $r$ for the QWZ model at the phase transition point, using a system that is 100 sites long and 500 sites wide. 
We choose the reference point $z= \frac{i}2$ that falls within the upper winding region in Fig.~3(b) of the main text. 
The spectrum of this Hamiltonian is plotted in Fig.~\ref{fig:finite_size}(a). 
There is a pair of modes (red) that are closest in energy to zero and are separated by a small gap from the rest of the spectrum. 
The inset shows that this pair of edge modes is strongly hybridized. 
Increasing the width $W$ of the scattering region reduces their hybridization energy, as shown in Fig.~\ref{fig:finite_size}(b), yielding a large localization length of $\xi\simeq 318$ sites. 
Therefore, if the scattering region at the topological phase transition point would be much wider than this localization length, its reflection matrix should exhibit the NHSE. 

\section{Goos-H\"anchen shift}

\begin{figure}[h!]
\includegraphics[width=\columnwidth]{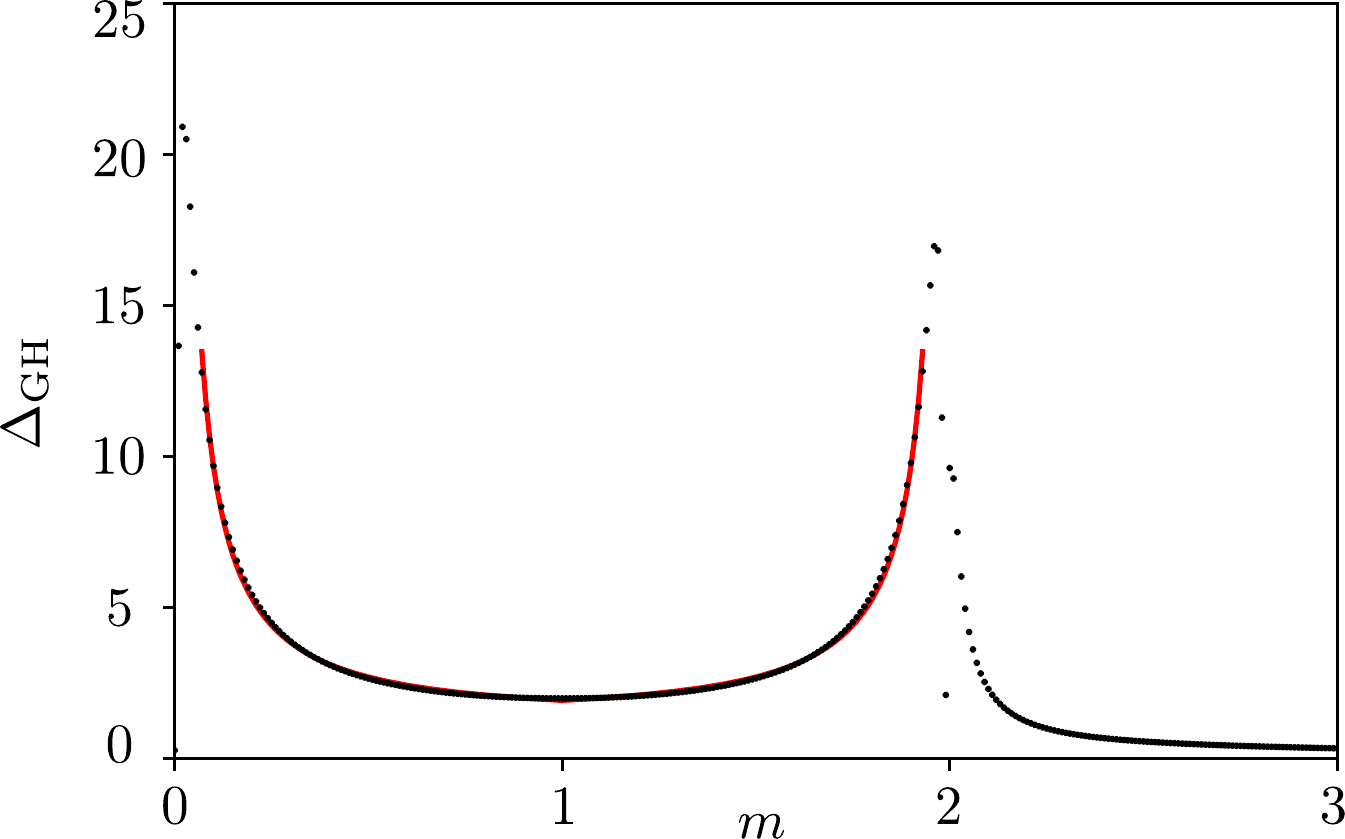}
\caption{
The dependence of the Goos-H\"anchen shift $\Delta_{\rm GH}$ on the parameter $m$. 
We calculate this shift for a Gaussian incoming wave packet with spread $s = 30$, while the scattering region is $80$ sites long and $200$ sites wide. 
Black points denote numerical data, and the red curve represents the fit function Eq.~\eqref{eq:GHfit} with parameters $a= 1.625$ and $b=0.159$.
}
\label{fig:gh}
\end{figure}

In Fig.~4 of the main text, we have shown that the incoming wave packet experiences a lateral shift upon reflection from a Chern insulator. 
This nonreciprocal Goos-H\"anchen (GH) shift occurs because the incoming wave packet couples to the chiral edge mode before being reflected. 
Here, we examine how this shift depends on the parameters of the scattering region. 
In our work, we modeled this scattering region with the QWZ model that has a single parameter $m$ that determines the Chern number of the bulk. 
Possible values of this topological invariant are $C=-1$ for $-2<m<0$, $C=1$ for $0<m<2$, and $C=0$ otherwise. 
In the following, we vary $m$ between $0$ and $3$, and calculate the magnitude of the lateral shift by subtracting the center of mass for the reflected wave packet from the center of mass of the incoming wave packet. 
The results are shown in Fig.~\ref{fig:gh}.

Firstly, we note that the lateral shift $\Delta_{\rm GH}$ vanishes in the trivial limit, $m\to \infty$, indicating its topological origin. 
In the range $m \in (0,2)$, we find that the lateral shift can be fitted with the function 
\begin{equation}
        \label{eq:GHfit}
        f(m) = \frac{a}{1-b^{1-|1-m|}},
\end{equation}
where $a$ and $b$ are fitting parameters. 
This function diverges at the topological phase transition points $m=0,2$. 
Such a behavior is reflected in the numerical results, where we see that the calculated lateral shifts have maximal values near these phase transition points. 
In these regions, finite size effects are very pronounced and are responsible for the mismatch between the fitting function and the numerical results.

To understand the dependence of the lateral shift $\Delta_{\rm GH}$ on the parameter $m$, we give a semi-classical reasoning. 
We think of the shift as an incoming mode that populates the chiral edge state and propagates along the edge of the Chern insulator. 
At every subsequent site in the transverse direction of the lead, this propagating mode has a finite probability $p$ to escape the edge state. 
Thus, the probability of escaping at the $n$-th site is given by the geometric distribution $P_n=p(1-p)^n$. 
The expectation value is $1/p$, which means that the GH shift will be proportional to this expectation value, $\Delta_{\rm GH}\propto 1/p$ (only proportional because not all of the incident mode couples to the edge mode).

In order to give an estimate for $p$, we use the following properties of the QWZ model. 
First, in Ref.~\cite{Peng2017}, it was shown analytically that the dispersion relation of the chiral edge mode in case of the QWZ model equals $\rm{sign}(m)\sin{k_y}$. 
Since this only depends on the sign of $m$ the velocity of the edge mode is $v=1$ at zero energy for the parameters considered in our work. 
Second, the localization length $\xi$ of the edge mode is inversely proportional to the bulk gap, and in the ribbon geometry the bulk gap is $E_{\rm gap} = 2 - 2|1-m|$. 

We can think of the electron as bouncing back and forth in the width $\xi$ of the edge state. We associate an escape probability $q$ to every bounce. 
The frequency of the bouncing is proportional to $1/\xi$ which is proportional to the bulk gap. 
The probability $p$ to escape during a given characteristic time (which depends on the edge state velocity, which for us is $v=1$) can be expressed as the cumulative distribution function of the exponential distribution $p=1-\mathrm{e}^{-\alpha E_{\rm gap}}$, where $\alpha$ is a constant parameter. 
Redefining the parameters as $b=\mathrm{e}^{-2\alpha}$ in the expectation value of the GH shift $\Delta_{\rm GH}=a/p$ we get Eq.~\eqref{eq:GHfit}.

In the region of small band gap, where $\xi$ and the GH shift are large enough compared to the lattice constant, we can give a more simple intuitive argument. 
Since the effect comes from the coupling of the incoming wave packet with the chiral edge mode, we can write the lateral shift as $\Delta_{\rm GH} = v t$, where $v$ is the group velocity of the chiral edge mode and $t$ is the dwell time of the wave packet, i.e., the time that it spends inside the chiral edge mode. 
The group velocity is constant and $t\propto\xi$, which means that for $m\to0$ the GH shift is $\Delta_{\rm GH} \propto 1/m$. 
This is consistent with Eq.~\eqref{eq:GHfit} in the limit of $m\to0$.

\bibliography{NHr}